# Experimental demonstration of cylindrical vector spatiotemporal optical vortex

Jian Chen,[1] Chenhao Wan,[1,2] Andy Chong,[3,4] and Qiwen Zhan[1,*]

[1]School of Optical-Electrical and Computer Engineering, University of Shanghai for Science and Technology, 200093 Shanghai, China
[2]School of Optical and Electronic Information, Huazhong University of Science and Technology, Wuhan, Hubei 430074, China
[3]Department of Physics, University of Dayton, 300 College Park, Dayton, Ohio 45469, USA
[4]Department of Electro-Optics and Photonics, University of Dayton, 300 College Park, Dayton, Ohio 45469, USA
*Corresponding authors: qwzhan@usst.edu.cn

**We experimentally generate cylindrically polarized wavepackets with transverse orbital angular momentum, demonstrating the coexistence of spatiotemporal optical vortex with spatial polarization singularity. The results in this paper extend the study of spatiotemporal wavepackets to a broader scope, paving the way for its applications in various areas such as light-matter interaction, optical tweezers, spatiotemporal spin-orbit angular momentum coupling, etc.**

Optical singularities are zero intensity points within a light field with undefined physical quantities, including scalar singularity arising from undefined phase and vector singularity arising from undefined state of polarization [1]. The commonly investigated scalar singularities are the optical vortex beams with orbital angular momentum (OAM) associated with a spatial spiral phase front [2,3]. Most previous researches on optical vortex beams dealt with longitudinal OAM, whose angular momentum is parallel to the propagation direction. Theoretical studies also reveal that the transverse OAM could be formed via introducing the temporal variations of phase [4], resulting in OAM perpendicular to the propagation direction of the beam. Recently, a linear method to experimentally generate spatiotemporal optical vortex (STOV) with transverse OAM in a controllable way has been reported [5] and the propagation of STOV in free space are studied with single-shot supercontinuum spectral interferometry [6]. Later wavepackets embedded with multiple spatiotemporal phase singularities carrying different transverse OAM has been demonstrated [7]. Wavepackets that contain both spatiotemporal phase singularity and spatial phase singularity has also been generated very recently [8]. Nevertheless, all these works are limited to the scalar singularities contained within the STOV wavepackets.

The most well known vector singularity is the cylindrical vector beams, including radial polarization and azimuthal polarization [9]. In this paper, we experimentally generate and characterize cylindrically polarized STOV with transverse OAM, demonstrating the coexistence of spatiotemporal OAM singularity with the spatial polarization singularity for the first time. The vectorial STOV may find applications in light-matter interaction, optical tweezers, spatiotemporal spin-orbit angular momentum coupling, etc.

The experimental setup to generate and characterize radially and azimuthally polarized STOVs are shown in Fig. 1. The laser source emits chirped pulses of about 3 ps duration as the incident wavepackets, whose central wavelength is 1030 nm. The incident pulse is split into two pulses by the non-polarizing beam splitter (NPBS) 1. One is directed to the pulse shaper consisting of a diffraction grating (G1), a cylindrical lens (CL) and one two-dimensional (2D) spatial light modulator (SLM) [10-12], in which the STOV is generated. Assuming that the optical field in the spatial frequency-frequency domain ($k_x - \omega$ domain) is given by $g_R(r)$ in the polar coordinates, where $r = \sqrt{k_x^2 + \omega^2}$ and $\theta = \tan^{-1}(\omega/k_x)$. The SLM located in the $k_x - \omega$ plane is used to apply spiral phase $e^{-il\theta}$ to the above field, after a 2D Fourier transformation, we can obtain the STOV in the x-t domain as the following [5]:

$$G(\rho,\phi) = \text{FT}\{g_R(r)e^{-il\theta}\} = 2\pi(-i)^l e^{-il\phi} H_l\{g_R(r)\}$$
$$H_l\{g_R(r)\} = \int_0^\infty r g_R(r) J_l(2\pi\rho r) dr \qquad (1)$$

where $\rho = \sqrt{x^2 + t^2}$ and $\phi = \tan^{-1}(x/t)$, $J_l(\cdot)$ is the Bessel function of the first kind, $l$ is the topological charge. Then, the STOV output from the pulse shaper is directed by Mirror 1 (M1) to pass through polarizer 1 (P1) and vortex wave plate (VWP). The transmission axis of P1 is along the horizontal direction, thus the transmitted beam is horizontally polarized, which will be converted into radially or azimuthally polarized STOV by the VWP subsequently. When the fast axis of the VWP is parallel with the polarization direction of the STOV, the transmitted wavepacket from the VWP is radially polarized, which can be expressed as

$$G_r(\rho,\phi,\beta) = \begin{pmatrix} \cos\beta \\ \sin\beta \end{pmatrix} G(\rho,\phi), \qquad (2)$$

where $\beta = \tan^{-1}(y/x)$ is the azimuthal angle in the spatial domain of the STOV. On the other hand, when the fast axis of the VWP is perpendicular to the polarization direction of the STOV, the transmitted wavepacket from the VWP is azimuthally polarized, as shown in the following:

$$G_a(\rho,\phi,\beta) = \begin{pmatrix} -\sin\beta \\ \cos\beta \end{pmatrix} G(\rho,\phi), \tag{3}$$

To characterize the generated vectorial STOVs, the other pulse divided by NPBS 1 is compressed to about 90 fs by a grating pair (G2 and G3) as the probe beam. Mirror 3 (M3) is used to adjust the time delay of the probe beam to make it interfere with the object beam at different time slice. Since both the radially and azimuthally polarized STOVs can be treated as a linear combination of vertically and horizontally polarized components, to reconstruct the three-dimensional (3D) spatiotemporal structure of each STOV, the probe beam should interfere with each of the polarization components at a series of time slices of the corresponding STOV. The polarizer 2 (P2) in the optical path of the probe beam is employed to make the probe beam horizontally polarized. And the half-wave plate (HWP) is adopted to rotate the polarization direction of the probe beam. Meanwhile, the intensity distributions of the object beam, probe beam and background noise also need to be measured individually in the characterization process [13-15]. Shutters 1 and 2 are utilized to select which beam will be measured by the CCD.

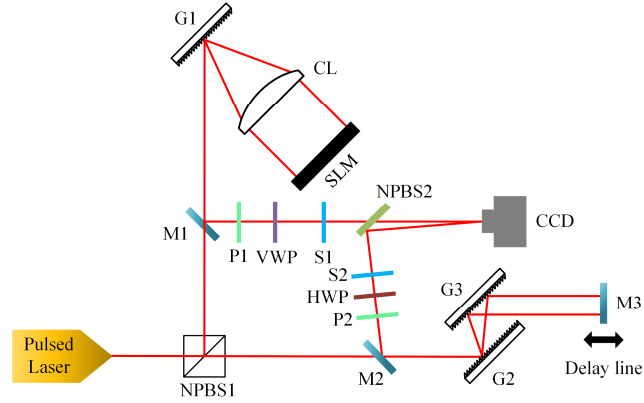

Fig. 1. Experimental setup for generating and characterizing vectorial spatiotemporal vortices. M: Mirror; G: Grating; CL: Cylindrical Lens; P: Polarizer; NPBS: Non-polarizing beam splitter; S: Shutter; VWP: Vortex wave plate.

In the first experiment, we applied a spiral phase pattern with topological charge of 1 to the SLM within the pulse shaper, and rotate the fast axis of the VWP to horizontal direction. To analyze the polarization distribution of the generated wavepacket, a linear polarizer together with a quarter wave plate are inserted between the NPBS2 and CCD to conduct the Stokes parameters measurement. The experimental results are shown in Fig. 2. The elevation angle has a nearly uniform distribution in the range of [-90°, 90°], and the ellipticity peaks around $0.01\pi$. Combing with the polarization map, we can conclude that the generated wavepacket is radially polarized.

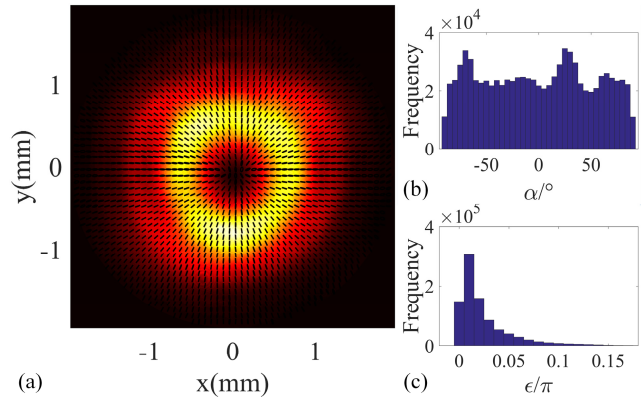

Fig. 2. Experimental results for the Stokes parameters measurement of radially polarized STOV. (a) Intensity distribution of the object beam with polarization map. Histograms of elevation angle (b) and ellipticity (c).

Subsequently, we open both shutters 1 and 2 to measure the 3D spatiotemporal structure of the generated wavepacket. The time delay is electronically controlled to step the probe beam throughout the object beam automatically. The fast axis of the HWP is firstly rotated to horizontal direction, thus the transmitted probe beam is horizontally polarized, which just interferes with the x-polarization component of the object beam at each time slice, as shown in Fig. 3(a). The interference patterns at several typical temporal locations are shown in Figs. 3(b)-3(e). We can see that there are no interference fringes near the vertical center line in all these patterns, which is due to the fact that the nearly vertical polarization within this area does not interference with the horizontally polarized probe. The evolution of the fringe patterns could be understood as the following. As the probe pulse is overlapped near the head of the spatiotemporal vortex, smooth continuous vertical fringes are observed due to a phase difference close to 0 or 2π between the upper and lower parts of the beam in the spatial domain, as shown in Fig. 3(b). As the probe pulse scans toward the center of the spatiotemporal vortex, the phase difference approaches to π due to the spatiotemporal vortex phase, the upper and lower fringes start to shift with respect to each other, as shown in Fig. 3(c). At the center of the spatiotemporal vortex, the upper and lower fringes are totally misaligned with each other because of the π phase difference between the corresponding two parts, as shown in Fig. 3(d). Note that from Fig. 3(a), the polarization in this part of the object beam points to the same direction. Just as the probe pulse passes the center, the phase difference becomes smaller than π, resulting the upper and lower fringes shifts in the opposite direction, as shown in Fig. 3(e). As the probe pulse advances more toward the end of the spatiotemporal vortex wavepacket, the fringes become vertically aligned again, which is similar to that in Fig. 3(b). The evolution of the fringe patterns is the consequence of the spiral phase in the spatiotemporal domain that has been experimentally measured in [5].

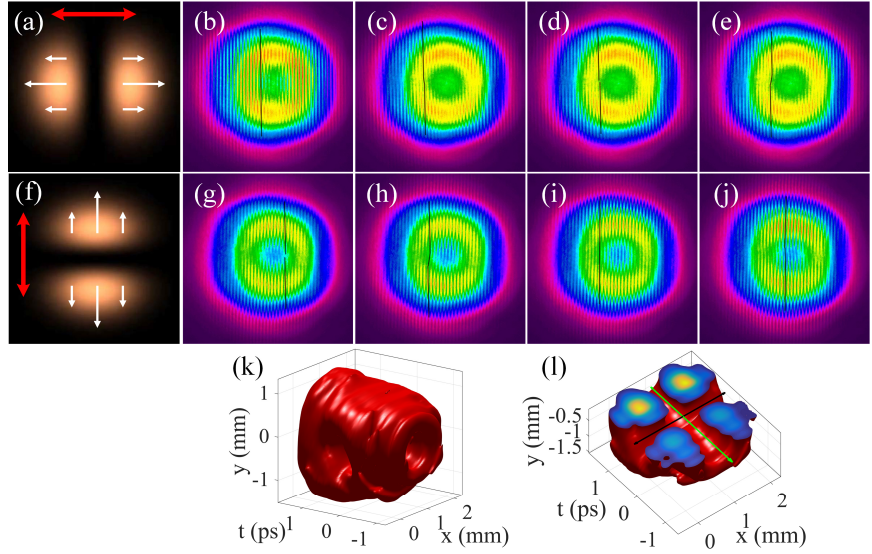

Fig. 3. Experimental results for characterizing radially polarized STOV. (a) Diagram of the pattern for the horizontally polarized component within a radially polarized object beam. (b-e) The interference patterns at several typical temporal locations when the probe beam is horizontally polarized. (f) Diagram of the pattern for the vertically polarized component within a radially polarized object beam. (g-j) The interference patterns at several typical temporal locations when the probe beam is vertically polarized. (k, l) The reconstructed 3D spatiotemporal wavepacket and its cut-through view. The red arrows in (a) and (f) indicate the polarization direction of the probe beam. The green and black lines in (l) indicate the traces of the polarization and OAM singularities, respectively.

Then the fast axis of the HWP is rotated by 45 degree to make the transmitted probe pulse be vertically polarized, which is used to scan through the wavepacket again. In this case, the probe beam will just interfere with the y-polarization component of the object beam, as shown in Fig. 3(f). The interference patterns at several typical temporal locations are shown in Figs. 3(g)-3(j). The differences from the former case are that there exist no fringes along the horizontal central axis, and the evolution of the fringe patterns also changes. Near the head or tail of the wavepacket, the upper and lower fringes totally misaligned with each other due to the π phase difference caused by the opposite polarization orientation within the upper-half and lower-half of the beam, while the phase difference contributed from the spatiotemporal phase pattern is near 0 or 2π. Towards the center of the spatiotemporal vortex, the fringes are aligned along the vertical direction. This is due to the contribution of π phase difference from the opposite polarization orientation and the π phase difference from the spatiotemporal vortex phase. Thus at this location the total phase difference across the upper-half and lower-half of the beam in Fig. 3(i) is 2π and the interference fringes will line up. Between the head to the center of the vortex or between the center to the tail of the vortex, the upper and lower fringes are misaligned but connected while the bending directions in these two transition regions are reversed.

Based on all the collected interference patterns corresponding to the two probe polarization, 3D spatiotemporal structure of the generated wavepacket can be reconstructed as shown in Fig. 3(k). To reveal singularity structures within the wavepacket, Fig. 3(l) demonstrate a cut-through view of the reconstructed wavepacket along the meridional plane. Two tunnels within the wavepacket marked by the green and black lines can be clearly seen. The transverse tunnel is caused by the phase singularity carried by the spatiotemporal spiral phase with topological charge of 1, while the longitudinal tunnel is caused by the polarization singularity carried by the radial polarization. Hereby, a radially polarized STOV with topological charge of 1 is generated.

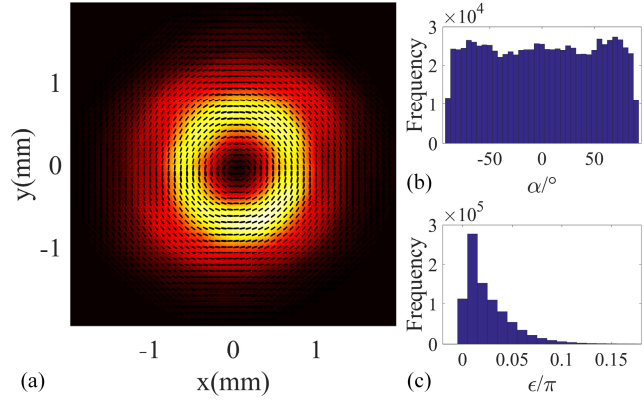

Fig. 4. Experimental results for the Stokes parameters measurement of the azimuthally polarized STOV. (a) Intensity distribution of the object beam with polarization map. Histograms of elevation angle (b) and ellipticity (c).

In the next experiment, we applied the same spiral phase pattern to the SLM, and rotate the fast axis of the VWP to vertical direction. The results of the Stokes parameters measurement for the transmitted wavepacket are shown in Fig. 4. From the polarization map and histograms of the elevation angle and ellipticity, we can confirm that the transmitted wavepacket is azimuthally polarized. The interference patterns between the probe beam and the object beam in several typical temporal slices are shown in Figs. 5(b)-5(e) and Figs. 5(g)-5(j). The patterns in the first row are obtained with a horizontally polarized probe pulse, while the patterns in the second row are collected with a vertically polarized probe pulse. Figures 5(a) and 5(f) give the diagram of the patterns for the horizontally and vertically polarized components of an azimuthally polarized object beam, respectively. The evolution of the interference fringe patterns can be explained similar to the radially polarized STOV case. There exist no fringes near the horizontal center line in Figs. 5(b)-5(e) and no fringes near the vertical center line in Figs. 5(g)-5(j), which are opposite to what is observed in Fig. 3. Likewise, the evolution of the fringe patterns are also contrary to those in Fig. 3, indicating the azimuthal polarization of generated wavepacket again. These differences in the fringe pattern behavior are due to the fact that the polarization pattern is orthogonal to the radially polarized STOV shown in Fig. 3. As shown in Figs. 5(k) and 5(l), the reconstructed 3D spatiotemporal structure and its cut-through view demonstrate that there are two types of optical singularity in the generated wavepacket. And the generated wavepacket is azimuthally polarized STOV with topological charge of 1.

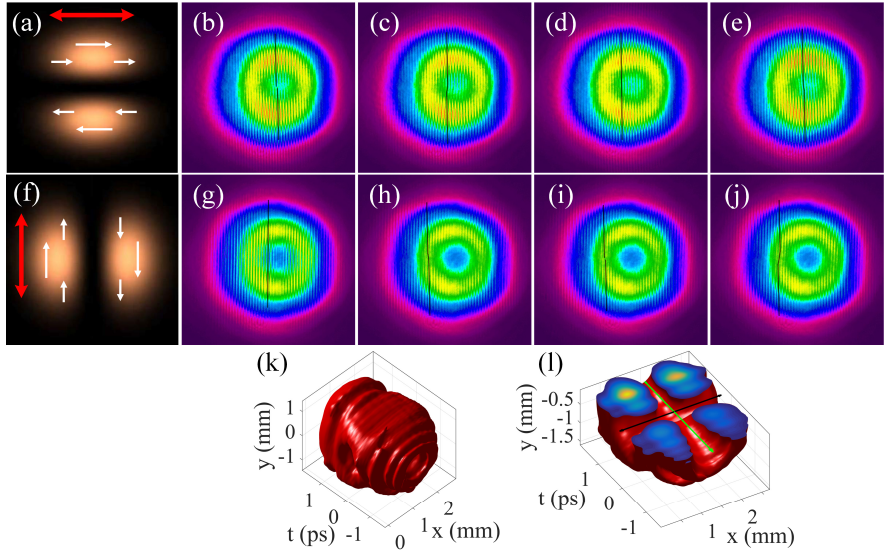

Fig. 5. Experimental results for characterizing azimuthally polarized STOV. (a) Diagram of the pattern for the horizontally polarized component within an azimuthally polarized object beam. (b-e) The interference patterns at several typical temporal locations when the probe beam is horizontally polarized. (f) Diagram of the pattern for the vertically polarized component within an azimuthally polarized object beam. (g-j) The interference patterns at several typical temporal locations when the probe beam is vertically polarized. (k, l) The reconstructed 3D spatiotemporal wavepacket and its cut-through view. The red arrows in (a) and (f) indicate the polarization direction of the probe beam. The green and black lines in (l) indicate the traces of the polarization and OAM singularities, respectively.

In summary, we experimentally generated and characterized radially and azimuthally polarized STOV with topological charge of 1. Both Stokes parameters measurement and 3D spatiotemporal measurement are conducted to characterize the generated cylindrically polarized

STOV. The results revealed that the spatiotemporal OAM singularity can coexist with the spatial polarization singularity. To our best knowledge, this is for the first time that optical wavepackets with both spatiotemporal singularity and polarization singularity have been reported. In the past two decades or so, there has been a tremendously increasing interests in the research and applications of spatial phase singularity such as OAM beams and vectorial singularity such as cylindrical vector beams. The discovery of spatiotemporal vortex and its controllable experimental creation and characterization, however, are very recent. The realization of optical wavepackets with both spatiotemporal phase singularity and spatial polarization singularity provides a completely new type of spatiotemporal optical field. We expect that further investigations into the properties of this new type of optical wavepackets and their interactions with matters will reveal more interesting findings and pave the way to a myriad of applications ranging from classical to quantum optics.

**Funding.** National Natural Science Foundation of China (92050202, 61805142, 61875245); Shanghai Science and Technology Committee (19060502500); Shanghai Natural Science Foundation (20ZR1437600).

**Disclosures.** The authors declare no conflicts of interest.